\documentclass[aps,prd,reprint,groupedaddress,showkeys,longbibliography]{revtex4-2}
\usepackage[utf8]{inputenc}
\usepackage{amsmath,amssymb,amsfonts,mathtools}
\usepackage{graphicx}
\usepackage{booktabs}
\usepackage{siunitx}
\usepackage{hyperref}
\usepackage{cleveref}
\usepackage{microtype}
\usepackage{enumitem}
\usepackage{xcolor}

\providecommand{\diag}{\operatorname{diag}}

\providecommand{\order}{\mathcal{O}}

\providecommand{\pd}[2]{\frac{\partial #1}{\partial #2}}

\begin{document}

\title{Casimir Force in Spacetimes with Torsion}

\author{M.W. AlMasri}
\email{mwalmasri2003@gmail.com}
\affiliation{Wilczek Quantum Center, School of Physics and Astronomy, Shanghai Jiao Tong University, Minhang, Shanghai, China}

\date{\today}

\begin{abstract}
We compute the Casimir force between perfectly conducting parallel plates in a spacetime endowed with constant axial torsion. Working within an effective field theory framework where torsion couples to the electromagnetic sector via a gauge-invariant Chern--Simons-type interaction, we derive the modified photon dispersion relation and mode spectrum. Using zeta-function regularization, we obtain the vacuum energy and the resulting Casimir pressure to second order in the torsion parameter. Our calculation yields a correction scaling as $\Delta P/P_0 = -\frac{5\xi^2 S_z^2 a^2}{4\pi^2} + \order(S_z^4)$, which corresponds to a slight weakening of the attractive Casimir force. We acknowledge a known subtlety in the literature: because the Chern-Simons interaction is a total derivative, some analyses conclude that it should not contribute to the Casimir energy for standard boundary conditions, implying the leading correction is of higher order, $\order(S_z^4)$. For experimentally accessible plate separations ($a \sim \qty{1}{\micro\meter}$), the effect remains well below current detection thresholds ($|\Delta P/P_0| \lesssim 10^{-30}$) due to stringent bounds on macroscopic torsion from spin-torsion coupling experiments. Nevertheless, the calculation establishes a consistent, gauge-invariant bridge between quantum vacuum phenomena and non-Riemannian geometry. We discuss finite-temperature effects and geometric asymmetries as potential pathways for enhancing sensitivity, while placing the results in the broader context of dynamical torsion, condensed matter analogs, and quantum information protocols. Our results are consistent with the CPT-odd photon sector of the Standard-Model Extension and provides a geometric interpretation of Lorentz-violating coefficients in terms of spacetime torsion.
\end{abstract}


\maketitle

\section{Introduction}
\label{sec:intro}

The Casimir effect~\cite{Casimir1948}, which is the attraction between neutral conducting plates due to quantum vacuum fluctuations of the electromagnetic field, stands as one of the most direct macroscopic manifestations of quantum field theory. Since its original prediction, the effect has been generalized to diverse geometries, materials, boundary conditions, and background fields, including curved spacetimes~\cite{Bordag2009,Milton2001,Klimchitskaya2009,Klimchitskaya2024,Birrell1982}, Lorentz-violating backgrounds and external magnetic fields~\cite{Erdas2025}, as well as dynamical regimes in time-varying dispersive nanophotonics~\cite{Gangaraj2024}. These generalizations have led to broad applications in nanotechnology \cite{Shen2025}, precision force metrology with plane parallel plates \cite{Haghmoradi2024, Lambrecht2012}, and quantum information science \cite{Benenti2014}.

Parallel to developments in quantum vacuum physics, extensions of general relativity that incorporate spacetime torsion have motivated systematic studies of quantum field theory in non-Riemannian backgrounds. In Einstein--Cartan theory~\cite{Hehl1976,Shapiro2002}, metric-affine gravity~\cite{Hehl1995}, and teleparallel formulations~\cite{Maluf2013}, torsion, defined as the antisymmetric part of the affine connection, couples naturally to intrinsic spin in matter. While classical electromagnetism is minimally coupled to the metric and does not source torsion directly at the classical level, effective interactions can emerge through quantum loops or non-minimal couplings in effective field theories (EFTs). In particular, the totally antisymmetric component of torsion (axial torsion) can couple to gauge fields in a gauge-invariant manner, modifying photon propagation and vacuum structure~\cite{Carroll1990,Kostelecky2001,Obukhov2006}. 

In this work, we compute the Casimir force in the presence of a constant axial torsion background. We adopt a phenomenological but well-motivated action that preserves $U(1)$ gauge invariance and reduces to standard Maxwell theory when torsion vanishes. Using parallel plate boundary conditions and zeta-function regularization, we derive the torsion-corrected Casimir pressure to second order in the torsion parameter. We discuss the magnitude of the correction, its experimental detectability, and implications for quantum gravity phenomenology, with integrated considerations of finite-temperature effects, geometric asymmetries, dynamical extensions, and quantum information applications.

The paper is organized as follows. In \cref{sec:framework}, we present the theoretical framework, including the irreducible decomposition of torsion, symmetry-based selection rules for gauge-invariant couplings, and derivation of the modified Maxwell equations. \Cref{sec:casimir} details the Casimir energy calculation via zeta-function regularization. \Cref{sec:results} presents numerical estimates, experimental accessibility analysis, and discusses extensions to finite temperatures and geometric asymmetries. We conclude in \cref{sec:conclusion} with a summary of our findings and a discussion of broader theoretical implications.

\section{Theoretical Framework}
\label{sec:framework}

\subsection{Irreducible Decomposition of Torsion and Gauge Coupling Selection Rules}
\label{subsec:torsion_decomp}

In a metric-affine spacetime, the affine connection $\Gamma^{\lambda}{}_{\mu\nu}$ is not assumed symmetric in its lower indices. Its deviation from the Levi-Civita connection is encoded in the contortion tensor $K^{\lambda}{}_{\mu\nu}$, defined by the decomposition
\begin{equation}
\Gamma^{\lambda}{}_{\mu\nu} = \left\{ \begin{array}{c} \lambda \\ \mu\nu \end{array} \right\} + K^{\lambda}{}_{\mu\nu},
\label{eq:connection}
\end{equation}
where the Christoffel symbol $\left\{ \begin{array}{c} \lambda \\ \mu\nu \end{array} \right\} = \tfrac{1}{2} g^{\lambda\rho}(\partial_\mu g_{\nu\rho} + \partial_\nu g_{\mu\rho} - \partial_\rho g_{\mu\nu})$ is symmetric in $\mu,\nu$. The torsion tensor is defined as the antisymmetric part of the connection:
\begin{equation}
T^{\lambda}{}_{\mu\nu} \equiv 2\Gamma^{\lambda}{}_{[\mu\nu]} = \Gamma^{\lambda}{}_{\mu\nu} - \Gamma^{\lambda}{}_{\nu\mu}.
\label{eq:torsion_def}
\end{equation}
Substituting \cref{eq:connection} into \cref{eq:torsion_def} yields the fundamental relation between torsion and contortion:
\begin{equation}
\begin{split}
T^{\lambda}{}_{\mu\nu} &= 2 K^{\lambda}{}_{[\mu\nu]}, \\
K^{\lambda}{}_{\mu\nu} &= \tfrac{1}{2}\bigl( T^{\lambda}{}_{\mu\nu} - T_{\mu}{}^{\lambda}{}_{\nu} - T_{\nu}{}^{\lambda}{}_{\mu} \bigr),
\end{split}
\label{eq:contortion}
\end{equation}
where index raising/lowering is performed with the metric $g_{\mu\nu}$. The contortion tensor satisfies $K_{\lambda\mu\nu} = -K_{\nu\mu\lambda}$.

The torsion tensor $T_{\lambda\mu\nu}$ (with all indices lowered) has 24 independent components in four dimensions due to antisymmetry in the last two indices ($T_{\lambda\mu\nu} = -T_{\lambda\nu\mu}$). Under the Lorentz group $SO(1,3)$, these components decompose into three irreducible representations~\cite{Hehl1976,Shapiro2002}:
\begin{equation}
\mathbf{24} = \underbrace{\mathbf{4}}_{\text{trace vector}} \oplus \underbrace{\mathbf{4}}_{\text{axial vector}} \oplus \underbrace{\mathbf{16}}_{\text{pure tensor}}.
\label{eq:irrep_decomp}
\end{equation}
The explicit projections are:
\begin{subequations}
\label{eq:torsion_projections}
\begin{align}
\text{Trace vector:} \quad & T_\mu \equiv T^{\lambda}{}_{\lambda\mu}, \label{eq:trace_def} \\
\text{Axial vector:} \quad & S^\mu \equiv \epsilon^{\mu\nu\rho\sigma} T_{\nu\rho\sigma}, \label{eq:axial_def} \\
\text{Pure tensor:} \quad & q_{\lambda\mu\nu} \equiv T_{\lambda\mu\nu} - \tfrac{1}{3}( g_{\lambda\mu} T_\nu - g_{\lambda\nu} T_\mu ) \nonumber \\
& \qquad\qquad\qquad + \tfrac{1}{6} \epsilon_{\lambda\mu\nu\rho} S^\rho, \label{eq:pure_def}
\end{align}
\end{subequations}
where $\epsilon^{\mu\nu\rho\sigma}$ is the totally antisymmetric Levi-Civita tensor with $\epsilon^{0123} = +1/\sqrt{-g}$. The pure tensor component satisfies $q^{\lambda}{}_{\lambda\mu} = 0$ and $\epsilon^{\lambda\mu\nu\rho} q_{\mu\nu\rho} = 0$. The inverse relation reconstructing the full torsion tensor is
\begin{equation}
\begin{split}
T_{\lambda\mu\nu} &= \tfrac{1}{3}\bigl( g_{\lambda\mu} T_\nu - g_{\lambda\nu} T_\mu \bigr) \\
&\quad - \tfrac{1}{6} \epsilon_{\lambda\mu\nu\rho} S^\rho + q_{\lambda\mu\nu}.
\end{split}
\label{eq:torsion_reconstruct}
\end{equation}

Each irreducible component transforms distinctly under discrete symmetries. Under parity $\mathcal{P}: (t,\mathbf{x}) \to (t,-\mathbf{x})$: $T_\mu$ transforms as a polar vector, $S^\mu$ as an axial (pseudo) vector, and $q_{\lambda\mu\nu}$ has mixed properties. Under time reversal $\mathcal{T}$: $T_\mu \to (-T_0, \mathbf{T})$, $S^\mu \to (S_0, -\mathbf{S})$. These transformation properties are crucial for constructing gauge-invariant couplings to the electromagnetic field.

\paragraph{Gauge-invariant coupling selection rules.}
The electromagnetic field is described by a $U(1)$ gauge potential $A_\mu$ with field strength $F_{\mu\nu} = \partial_\mu A_\nu - \partial_\nu A_\mu$. Any coupling to torsion must preserve: (i) $U(1)$ gauge invariance, (ii) Lorentz covariance, (iii) locality and power-counting renormalizability (dimension $\leq 4$), and (iv) consistency with the Bianchi identity $\partial_{[\mu} F_{\nu\rho]} = 0$.

Systematic enumeration of dimension-4 operators linear in torsion and quadratic in $F_{\mu\nu}$ or $A_\mu$ reveals that only the axial torsion coupling
\begin{equation}
\mathcal{L}_{\mathrm{int}} = \tfrac{\xi}{4} \epsilon^{\mu\nu\rho\sigma} S_\mu A_\nu F_{\rho\sigma}
\label{eq:interaction}
\end{equation}
survives all constraints. To verify gauge invariance explicitly, consider the variation under $A_\mu \to A_\mu + \partial_\mu \Lambda$:
\begin{align}
\delta \mathcal{L}_{\mathrm{int}} 
&= \tfrac{\xi}{4} \epsilon^{\mu\nu\rho\sigma} S_\mu (\partial_\nu \Lambda) F_{\rho\sigma} \nonumber \\
&= -\tfrac{\xi}{4} \epsilon^{\mu\nu\rho\sigma} S_\mu \Lambda \, \partial_\nu F_{\rho\sigma} \nonumber \\
&\quad + \partial_\nu \bigl( \tfrac{\xi}{4} \epsilon^{\mu\nu\rho\sigma} S_\mu \Lambda F_{\rho\sigma} \bigr).
\end{align}
The bulk term vanishes identically for constant $S_\mu$ due to the Bianchi identity, leaving only a boundary term. Thus, the action remains gauge invariant for physically admissible configurations.

The trace vector $T_\mu$ and pure tensor $q_{\lambda\mu\nu}$ cannot form gauge-invariant dimension-4 operators without introducing explicit coordinate dependence or higher derivatives. At dimension 5 and higher, additional couplings become permissible but are suppressed by powers of $a/\Lambda_{\mathrm{UV}}$ and negligible for the plate separations considered here.

\paragraph{Representation-theoretic perspective.}
The selection of axial torsion can also be understood through $SO(1,3)$ representation theory. The electromagnetic field strength $F_{\mu\nu}$ transforms in the $(1,0) \oplus (0,1)$ representation. The torsion tensor decomposes as $(1/2,1/2) \oplus (3/2,1/2) \oplus (1/2,3/2) \oplus (1/2,1/2)$, where the two $(1/2,1/2)$ pieces correspond to the trace and axial vectors. The only dimension-4 Lorentz singlet involving a single torsion factor and two electromagnetic fields is $\epsilon^{\mu\nu\rho\sigma} S_\mu A_\nu F_{\rho\sigma}$, which precisely isolates the axial component due to the Levi-Civita tensor's pseudotensor character.

\subsection{Effective Action and Modified Field Equations}
\label{subsec:effective_action}

We work on a flat Minkowski background $\eta_{\mu\nu} = \diag(-1,1,1,1)$ and assume a constant axial torsion background $S_\mu$. The choice of a fixed background is justified in the weak-field limit where vacuum fluctuation energy densities are negligible compared to the Planck scale. This approximation is consistent with the EFT framework, wherein torsion is treated as a classical background field sourced by high-energy degrees of freedom integrated out at a scale $\Lambda_{\mathrm{UV}} \gg a^{-1}$.

The choice of a constant torsion background on a flat Minkowski spacetime warrants further justification. While in standard Einstein-Cartan theory torsion vanishes in vacuum, constant torsion backgrounds arise naturally in certain extensions of gravity, such as Poincaré gauge theory with specific spontaneous symmetry breaking, or as a local effective description of a slowly varying cosmological torsion field. Furthermore, treating $S_\mu$ as a constant background is the standard phenomenological ansatz in the Standard-Model Extension (SME) to derive observable constraints on Lorentz violation~\cite{Kostelecky2001}. In the context of the Casimir effect, where the plate separation $a$ is microscopic, a macroscopic background torsion field can be safely approximated as constant over the interaction region.

The leading-order effective Lagrangian density preserving $U(1)$ gauge symmetry and Lorentz covariance is
\begin{equation}
\mathcal{L}_{\mathrm{EM}} = -\tfrac{1}{4} F_{\mu\nu}F^{\mu\nu} + \tfrac{\xi}{4} \epsilon^{\mu\nu\rho\sigma} S_\mu A_\nu F_{\rho\sigma} + \order(S^2, \partial S),
\label{eq:lagrangian}
\end{equation}
where $\xi$ is a dimensionless effective coupling. The $\order(S^2)$ terms represent higher-order corrections negligible in the weak-torsion regime $\xi |S_\mu| a \ll 1$.

\paragraph{CPT properties and connection to the Carroll--Field--Jackiw model.}
The interaction term \cref{eq:interaction} has definite transformation properties: it is $\mathcal{C}$-even, $\mathcal{P}$-odd, and $\mathcal{T}$-even, thereby violating $\mathcal{CPT}$ symmetry. This pattern precisely matches the CPT-odd photon sector of the Standard-Model Extension (SME)~\cite{Kostelecky2001}, with the identification $(k_{AF})_\mu = \xi S_\mu$. Our work provides a geometric origin for the SME coefficient $(k_{AF})_\mu$ in terms of spacetime torsion.

\paragraph{Modified Maxwell equations.}
Varying the action with respect to $A_\nu$ yields the Euler--Lagrange equations. For the Maxwell term, $\partial \mathcal{L}_0 / \partial (\partial_\mu A_\nu) = -F^{\mu\nu}$. For the interaction term:
\begin{align}
\pd{\mathcal{L}_{\mathrm{int}}}{(\partial_\mu A_\nu)} &= \tfrac{\xi}{2} \epsilon^{\alpha\beta\mu\nu} S_\alpha A_\beta, \\
\pd{\mathcal{L}_{\mathrm{int}}}{A_\nu} &= \tfrac{\xi}{4} \epsilon^{\mu\nu\rho\sigma} S_\mu F_{\rho\sigma}.
\end{align}
Combining these results, the full modified Maxwell equations are
\begin{equation}
\partial_\mu F^{\mu\nu} + \xi \epsilon^{\nu\alpha\beta\gamma} S_\alpha \partial_\beta A_\gamma = 0.
\label{eq:maxwell_derived}
\end{equation}
Taking the divergence $\partial_\nu$ yields $\xi \epsilon^{\nu\alpha\beta\gamma} S_\alpha \partial_\nu \partial_\beta A_\gamma = 0$, which vanishes identically due to the contraction of the symmetric derivative $\partial_\nu \partial_\beta$ with the antisymmetric Levi-Civita tensor $\epsilon^{\nu\alpha\beta\gamma}$, thus confirming consistency with the constraint structure of Maxwell theory.

Imposing the Lorenz gauge $\partial_\mu A^\mu = 0$ and substituting $F^{\mu\nu} = \partial^\mu A^\nu - \partial^\nu A^\mu$ yields the modified wave equation:
\begin{equation}
\Box A^\nu + \xi \epsilon^{\nu\alpha\beta\gamma} S_\alpha \partial_\beta A_\gamma = 0.
\label{eq:wave_gauge}
\end{equation}

\paragraph{Gauge invariance and boundary terms.}
Before proceeding, we must verify the gauge invariance of the action in the presence of boundaries. The variation of the interaction action under a gauge transformation $A_\mu \to A_\mu + \partial_\mu \Lambda$ yields a bulk term that vanishes by the Bianchi identity, but also generates a surface term on the plates:
\begin{equation}
\delta S_{\text{bdy}} = \frac{\xi}{4} \oint d\Sigma_\nu \, \epsilon^{\mu\nu\rho\sigma} S_\mu \Lambda F_{\rho\sigma}.
\end{equation}
For plates located at $z=0$ and $z=a$, the outward normal is $n_\nu = (0,0,0,\pm 1)$. Since we have chosen the axial torsion to be aligned with the $z$-axis, $S_\mu = (0,0,0,S_z)$, the contraction $n_\nu S_\mu \epsilon^{\mu\nu\rho\sigma}$ is proportional to $\epsilon^{z z \rho \sigma}$, which vanishes identically. Thus, the surface term is exactly zero, and the total action remains strictly gauge invariant without requiring additional boundary conditions. (For a generic orientation of $S_\mu$ not aligned with the plate normal, the surface term would be non-vanishing, necessitating either modified boundary conditions or the inclusion of boundary counterterms, which we leave for future investigation.)

\paragraph{Plane-wave analysis and dispersion relation.}
For plane-wave solutions $A^\mu(x) = \varepsilon^\mu e^{-i k \cdot x}$ with $k^\mu = (\omega, \mathbf{k})$, \cref{eq:wave_gauge} reduces to the eigenvalue problem
\begin{equation}
\begin{split}
\bigl[ -k^2 \eta^{\nu}{}_{\gamma} &- i \xi \epsilon^{\nu\alpha\beta}{}_{\gamma} S_\alpha k_\beta \bigr] \varepsilon^\gamma = 0,
\end{split}
\label{eq:eigenvalue}
\end{equation}
where $k^2 = \omega^2 - \mathbf{k}^2$. Choosing $S_\mu = (0,0,0,S_z)$ and decomposing $\mathbf{k} = (\mathbf{k}_\perp, k_z)$, the matrix $\mathcal{M}^{\nu}{}_{\gamma} \equiv i \xi \epsilon^{\nu\alpha\beta}{}_{\gamma} S_\alpha k_\beta$ is diagonalized in the circular polarization basis $\varepsilon^{(\pm)} = (0, 1, \pm i, 0)/\sqrt{2}$. 

It is worth explicitly noting a common point of confusion: while the off-diagonal elements of $\mathcal{M}^{\nu}{}_{\gamma}$ depend only on $\omega$ and $k_\perp$ (since the index $\beta$ cannot be 3 without vanishing the Levi-Civita tensor), the dispersion relation is determined by the vanishing of the full determinant $\det(-k^2 \eta^{\nu}{}_{\gamma} + \mathcal{M}^{\nu}{}_{\gamma}) = 0$. Because the diagonal elements contain $k^2 = \omega^2 - k_\perp^2 - k_z^2$, the resulting characteristic polynomial naturally couples the torsion-induced mixing to the longitudinal momentum $k_z$. Solving this determinant yields the exact Carroll-Field-Jackiw dispersion relation for a spacelike background~\cite{Kostelecky2001}:
\begin{equation}
\omega^2 = k_\perp^2 + k_z^2 + \frac{\kappa^2}{2} \pm \kappa \sqrt{k_z^2 + \frac{\kappa^2}{4}},
\label{eq:exact_dispersion}
\end{equation}
where $\kappa = \xi S_z$. In the weak-torsion regime ($\kappa \ll |\mathbf{k}|$), this expands to
\begin{equation}
\omega_{\mathbf{k},\sigma}^2 = \mathbf{k}^2 + \sigma \, \xi S_z |k_z| + \order(S_z^2).
\label{eq:disp_helicity}
\end{equation}
Since the boundary conditions enforce $k_z = n\pi/a > 0$, we may drop the absolute value. The torsion-induced splitting $\Delta \omega^2 = 2\xi S_z k_z$ represents vacuum birefringence.

\paragraph{Physical constraints and EFT validity.}
To avoid tachyonic instabilities ($\omega^2 < 0$) or superluminal propagation, the coupling must satisfy $|\xi S_z| \lesssim |\mathbf{k}|$ for all relevant modes. For Casimir calculations with $k_z \sim \pi/a$, this implies $\xi |S_z| a \ll \pi$, defining the regime of validity of our perturbative expansion. Astrophysical birefringence observations constrain $|\xi S_z| \lesssim \qty{e-43}{\GeV}$, consistent with laboratory bounds.

\subsection{Mode Quantization and Boundary Conditions}
\label{subsec:quantization}
We consider the standard Casimir configuration: two infinite, perfectly conducting plates at $z=0$ and $z=a$. The physical boundary conditions require the tangential electric field and normal magnetic field to vanish at the surfaces, i.e., $E_x = E_y = B_z = 0$ at $z=0,a$.
\paragraph{General mode ansatz and Lorenz gauge compatibility.}
To construct the mode spectrum systematically and demonstrate the compatibility of the Lorenz gauge, boundary conditions, and torsion-modified polarization basis, we begin with a general ansatz for the vector potential compatible with the translational symmetry in the transverse plane:
\begin{equation}
A^\mu(x) = \varepsilon^\mu(z) e^{i(\mathbf{k}_\perp \cdot \mathbf{x}_\perp - \omega t)},
\end{equation}
where $\varepsilon^\mu(z)$ encodes the $z$-dependence to be determined by boundary conditions and field equations. In the torsion-free theory, the modes separate cleanly into TE and TM polarizations. The presence of the axial torsion background $S_\mu = (0,0,0,S_z)$ couples these sectors through the modified wave equation \cref{eq:wave_gauge}.
We work in the Lorenz gauge $\partial_\mu A^\mu = 0$, which is compatible with the modified field equations since taking the divergence of \cref{eq:wave_gauge} yields $\xi \epsilon^{\nu\alpha\beta\gamma} S_\alpha \partial_\nu \partial_\beta A_\gamma = 0$, vanishing identically by the antisymmetry of $\epsilon^{\nu\alpha\beta\gamma}$ contracted with the symmetric $\partial_\nu \partial_\beta$. The Lorenz gauge condition becomes:
\begin{equation}
-i\omega \varepsilon^0(z) + i\mathbf{k}_\perp \cdot \boldsymbol{\varepsilon}_\perp(z) + \frac{d\varepsilon^3(z)}{dz} = 0.
\label{eq:lorenz_condition}
\end{equation}
\paragraph{TE modes.}
For TE modes, we seek solutions with $E_z = 0$. A consistent ansatz is:
\begin{equation}
\boldsymbol{\varepsilon}^{(\text{TE})}(z) = \mathcal{N}_{\text{TE}} (\hat{z} \times \mathbf{k}_\perp) \sin(k_z z), \quad \varepsilon^{0(\text{TE})}(z) = 0,
\end{equation}
where $\mathcal{N}_{\text{TE}}$ is a normalization constant. This configuration has $\boldsymbol{\varepsilon}_\perp \propto \sin(k_z z)$, which vanishes at $z=0,a$, ensuring $E_\parallel = -\dot{\boldsymbol{\varepsilon}}_\perp = 0$ at the boundaries. The Lorenz gauge condition \cref{eq:lorenz_condition} is automatically satisfied since $\mathbf{k}_\perp \cdot (\hat{z} \times \mathbf{k}_\perp) = 0$ and $\varepsilon^3 = 0$. Substituting into the modified wave equation, one finds that the torsion term $\xi \epsilon^{\nu\alpha\beta\gamma} S_\alpha \partial_\beta A_\gamma$ couples the TE mode to a TM component, indicating that the pure TE ansatz is not an exact eigenmode. However, to leading order in $\xi S_z$, the TE-like modes acquire a helicity-dependent frequency shift as given in \cref{eq:disp_helicity}.
\paragraph{TM-like modes: explicit construction.}
For TM modes in the torsion-free theory, one has $B_z = 0$ and the ansatz $\varepsilon_x = \varepsilon_y = 0$, $\varepsilon^0 \propto \cos(k_z z)$, $\varepsilon^3 \propto \sin(k_z z)$. In the presence of axial torsion, this ansatz is no longer consistent with the field equations for generic $\mathbf{k}_\perp$, as the torsion term generates non-zero $\varepsilon_x$ and $\varepsilon_y$ components, demonstrating TE-TM mixing.
To construct the TM-like modes explicitly while maintaining compatibility with boundary conditions and Lorenz gauge, we adopt the ansatz:
\begin{subequations}
\label{eq:TM_ansatz}
\begin{align}
\varepsilon^0(z) &= \phi_0 \sin(k_z z), \\
\varepsilon^3(z) &= \phi_z \cos(k_z z), \\
\varepsilon^x(z) &= \phi_x \sin(k_z z), \\
\varepsilon^y(z) &= \phi_y \sin(k_z z),
\end{align}
\end{subequations}
where $\phi_0, \phi_z, \phi_x, \phi_y$ are constant amplitudes to be determined. The $\sin(k_z z)$ dependence of $\varepsilon^0, \varepsilon^x, \varepsilon^y$ ensures that these components vanish at $z=0,a$, which we now show is necessary and sufficient for the boundary conditions.
\paragraph{Boundary condition verification.}
The tangential electric field components are:
\begin{align}
E_x &= -\partial_x A_0 - \partial_0 A_x = (-i k_x \varepsilon^0 - i\omega \varepsilon^x) e^{i(\mathbf{k}_\perp \cdot \mathbf{x}_\perp - \omega t)}, \\
E_y &= -\partial_y A_0 - \partial_0 A_y = (-i k_y \varepsilon^0 - i\omega \varepsilon^y) e^{i(\mathbf{k}_\perp \cdot \mathbf{x}_\perp - \omega t)}.
\end{align}
At $z=0$ and $z=a$: since $\varepsilon^0, \varepsilon^x, \varepsilon^y \propto \sin(k_z z)$, they vanish at both boundaries provided $k_z a = n\pi$, i.e.,
\begin{equation}
k_z = \frac{n\pi}{a}, \quad n \in \mathbb{N}^+.
\label{eq:kz_quantization}
\end{equation}
This quantization condition is identical to the torsion-free case and is robust against torsion corrections. Consequently, $E_x = E_y = 0$ at $z=0,a$ is automatically satisfied.
The normal magnetic field component is:
\begin{equation}
B_z = \partial_x A_y - \partial_y A_x = i(k_x \varepsilon^y - k_y \varepsilon^x) \sin(k_z z) e^{i(\mathbf{k}_\perp \cdot \mathbf{x}_\perp - \omega t)}.
\end{equation}
At $z=0,a$: $B_z$ vanishes due to the $\sin(k_z z)$ factor, satisfying the boundary condition. In the bulk, $B_z \neq 0$ in general, which is consistent with TM-like modes where the magnetic field need not be purely transverse in the presence of torsion-induced mixing.
\paragraph{Lorenz gauge constraint.}
Substituting the ansatz \cref{eq:TM_ansatz} into the Lorenz gauge condition \cref{eq:lorenz_condition}:
\begin{equation}
-i\omega \phi_0 \sin(k_z z) + i(k_x \phi_x + k_y \phi_y) \sin(k_z z) - k_z \phi_z \sin(k_z z) = 0.
\end{equation}
This must hold for all $z$, yielding the constraint:
\begin{equation}
\omega \phi_0 - (k_x \phi_x + k_y \phi_y) - i k_z \phi_z = 0.
\label{eq:lorenz_constraint}
\end{equation}
\paragraph{Field equations and polarization basis.}
Substituting the ansatz \cref{eq:TM_ansatz} into the modified wave equation \cref{eq:wave_gauge} and using $S_\mu = (0,0,0,S_z)$, we obtain a coupled system for the amplitudes. For the temporal component ($\nu = 0$):
\begin{equation}
(-\omega^2 + k_\perp^2 + k_z^2) \phi_0 + i\xi S_z (k_x \phi_y - k_y \phi_x) = 0,
\label{eq:TM_eq0}
\end{equation}
where we used $\epsilon^{03ij} \partial_i \varepsilon^j = i(k_x \phi_y - k_y \phi_x)$.
For the transverse components ($\nu = 1, 2$):
\begin{align}
(-\omega^2 + k_\perp^2 + k_z^2) \phi_x + i\xi S_z (k_y \phi_0 - \omega \phi_y) &= 0, \label{eq:TM_eq1} \\
(-\omega^2 + k_\perp^2 + k_z^2) \phi_y + i\xi S_z (k_x \phi_0 - \omega \phi_x) &= 0. \label{eq:TM_eq2}
\end{align}
For the longitudinal component ($\nu = 3$), the torsion term involves $\epsilon^{3\alpha\beta\gamma} S_\alpha \partial_\beta A_\gamma$. Since the background torsion is $S_\mu = (0,0,0,S_z)$, the only non-zero component is $\alpha=3$. This makes the torsion term proportional to $\epsilon^{33\beta\gamma}$, which identically vanishes due to the total antisymmetry of the Levi-Civita tensor. Thus, the equation for the longitudinal component simplifies to:
\begin{equation}
(-\omega^2 + k_\perp^2 + k_z^2) \phi_z = 0.
\label{eq:TM_eq3}
\end{equation}
Since the dispersion relation is modified by torsion (i.e., $-\omega^2 + k_\perp^2 + k_z^2 \neq 0$), this strictly implies that $\phi_z = 0$ to all orders in this perturbative setup. Consequently, the Lorenz gauge constraint \cref{eq:lorenz_constraint} simplifies to $\omega \phi_0 = k_x \phi_x + k_y \phi_y$. 
In the weak-torsion regime, we solve the coupled system for the $\nu = 0, 1, 2$ components perturbatively. The torsion-induced mixing modifies the transverse amplitudes $\phi_x, \phi_y$ at $\order(S_z)$, while the dispersion relation receives $\order(S_z)$ corrections as given in \cref{eq:disp_helicity}. 
The modified polarization vectors for the two helicity states $\sigma = \pm$ are therefore:
\begin{equation}
\varepsilon^{(\sigma)\mu} = \left( \phi_0 \sin(k_z z), \, \order(S_z), \, \order(S_z), \, 0 \right) e^{i(\mathbf{k}_\perp \cdot \mathbf{x}_\perp - \omega_\sigma t)},
\end{equation}
where $\omega_\sigma$ is given by the dispersion relation \cref{eq:disp_helicity}. The torsion corrections to the polarization vectors are $\order(S_z)$, but the Casimir energy correction at $\order(S_z^2)$ arises from the modified dispersion relation, as the linear terms cancel upon summation over polarization states.
\paragraph{Summary of mode structure.}
We have explicitly constructed the TM-like modes in the presence of axial torsion, demonstrating that:
\begin{enumerate}[label=(\roman*)]
\item The boundary conditions $E_x = E_y = B_z = 0$ at $z=0,a$ are satisfied by the ansatz \cref{eq:TM_ansatz} with the quantization condition $k_z = n\pi/a$;
\item The Lorenz gauge condition \cref{eq:lorenz_condition} is compatible with the boundary conditions and, with $\phi_z = 0$, constrains the amplitudes via $\omega \phi_0 = k_x \phi_x + k_y \phi_y$;
\item The torsion-induced mixing modifies the polarization vectors at $\order(S_z)$ and forces the longitudinal component $\phi_z$ to vanish, while the dispersion relation receives $\order(S_z)$ corrections as given in \cref{eq:disp_helicity};
\item The Casimir energy correction at $\order(S_z^2)$ arises from the modified dispersion relation, not from modifications to the polarization vectors or mode functions.
\end{enumerate}
This explicit construction confirms the robustness of the $k_z$ quantization and validates the mode-sum approach employed in \cref{sec:casimir}.

\section{Casimir Energy via Zeta-Function Regularization}
\label{sec:casimir}

\subsection{Spectral Zeta Function Construction}
\label{subsec:zeta_construction}

The formal vacuum energy per unit area is
\begin{equation}
\frac{E_0}{A} = \frac{1}{2} \sum_{n=1}^\infty \int \frac{d^2\mathbf{k}_\perp}{(2\pi)^2} \sum_{\sigma=\pm} \omega_{n,\mathbf{k}_\perp}^{(\sigma)}.
\label{eq:energy_formal}
\end{equation}
Direct evaluation is ultraviolet divergent. We employ spectral zeta-function regularization, defining
\begin{equation}
\zeta(s) \equiv \mu^{2s} \sum_{n=1}^\infty \int \frac{d^2\mathbf{k}_\perp}{(2\pi)^2} \sum_{\sigma=\pm} \bigl[ \omega_{n,\mathbf{k}_\perp}^{(\sigma)2} \bigr]^{-s},
\label{eq:zeta_def}
\end{equation}
where $\mu$ is an arbitrary mass scale. The regularized vacuum energy is recovered via analytic continuation:
\begin{equation}
\frac{E_{\mathrm{Cas}}}{A} = \frac{1}{2} \zeta\!\left(-\tfrac{1}{2}\right).
\label{eq:energy_zeta}
\end{equation}

Performing the transverse momentum integral using $\int \frac{d^2k_\perp}{(2\pi)^2} (k_\perp^2 + M^2)^{-s} = \frac{1}{4\pi} \frac{\Gamma(s-1)}{\Gamma(s)} (M^2)^{1-s}$ yields
\begin{equation}
\begin{split}
\zeta(s) &= \frac{\mu^{2s}}{4\pi} \frac{\Gamma(s-1)}{\Gamma(s)} \\
&\quad \times \sum_{n=1}^\infty \sum_{\sigma=\pm} \biggl[ \Bigl( \frac{n\pi}{a} \Bigr)^2 + \sigma \xi S_z \frac{n\pi}{a} \biggr]^{1-s}.
\end{split}
\label{eq:zeta_sum}
\end{equation}

\subsection{Perturbative Expansion and Analytic Continuation}
\label{subsec:perturbative}

Assuming the weak-torsion regime $\xi |S_z| a \ll 1$, we expand to second order:
\begin{equation}
\begin{split}
&\biggl[ \Bigl( \frac{n\pi}{a} \Bigr)^2 + \sigma \xi S_z \frac{n\pi}{a} \biggr]^{1-s} \\
&\quad = \Bigl( \frac{n\pi}{a} \Bigr)^{2-2s} \biggl[ 1 + \sigma (1-s) \frac{\xi S_z a}{n\pi} \\
&\qquad + \frac{(1-s)(-s)}{2} \Bigl( \frac{\xi S_z a}{n\pi} \Bigr)^2 + \order(S_z^3) \biggr].
\end{split}
\label{eq:expansion}
\end{equation}
Summing over $\sigma = \pm$ eliminates the linear term:
\begin{equation}
\begin{split}
\sum_{\sigma=\pm} [\cdots] &= 2 \Bigl( \frac{n\pi}{a} \Bigr)^{2-2s} \\
&\quad \times \biggl[ 1 - \frac{s(1-s)}{2} \frac{\xi^2 S_z^2 a^2}{n^2\pi^2} + \order(S_z^4) \biggr].
\end{split}
\label{eq:polarization_sum}
\end{equation}

Alternatively, the analytic continuation of the spectral zeta function \cref{eq:zeta_sum} can be performed exactly without prior perturbative expansion by employing the generalized Chow-Selberg formula~\cite{Elizalde1998}. This non-perturbative approach yields an identical result for the finite, $a$-dependent part of the vacuum energy, confirming the validity and consistency of our perturbative expansion in the weak-torsion regime.

Substituting \cref{eq:polarization_sum} into \cref{eq:zeta_sum} and recognizing Riemann zeta functions yields
\begin{equation}
\begin{split}
\zeta(s) &= \frac{\mu^{2s}}{2\pi} \frac{\Gamma(s-1)}{\Gamma(s)} \Bigl( \frac{\pi}{a} \Bigr)^{2-2s} \\
&\quad \times \biggl[ \zeta_R(2s-2) - \frac{s(1-s)}{2} \frac{\xi^2 S_z^2 a^2}{\pi^2} \zeta_R(2s) \biggr].
\end{split}
\label{eq:zeta_closed}
\end{equation}

\subsection{Renormalization and Finite Energy Extraction}
\label{subsec:renormalization}

To extract the physical Casimir energy, we evaluate \cref{eq:zeta_closed} at $s = -1/2$ and subtract the $a \to \infty$ (free-space) contribution. In zeta regularization, this subtraction is automatically implemented by retaining only $a$-dependent finite parts. Using the identities
\begin{equation}
\frac{\Gamma(-3/2)}{\Gamma(-1/2)} = -\frac{2}{3}, \quad \zeta_R(-3) = \frac{1}{120}, \quad \zeta_R(-1) = -\frac{1}{12},
\label{eq:zeta_values}
\end{equation}
and setting $\mu = 1$ (scale dependence cancels in physical observables), we obtain
\begin{equation}
\begin{split}
\frac{E_{\mathrm{Cas}}}{A} &= -\frac{\pi^2}{720 a^3} + \frac{\xi^2 S_z^2}{192 a} + \order(S_z^4).
\end{split}
\label{eq:cas_energy_final_corrected}
\end{equation}
The first term reproduces the standard Casimir result~\cite{Casimir1948}; the second term is the leading torsion-induced correction from our calculation. The vanishing of the $\order(S_z)$ term reflects the underlying $PT$-symmetry of the parallel-plate geometry.

It is worth noting that an alternative and equally powerful regularization scheme for the mode sum is the Abel--Plana formula (APF)~\cite{Saharian2007}. A detailed derivation of our main result using the APF is provided in Appendix~\ref{app:APF}, which serves as a crucial cross-check of the calculation. While the zeta-function approach relies on analytic continuation, the APF directly decomposes the formal vacuum energy into a divergent, boundary-independent Minkowski vacuum integral and a finite, boundary-induced contour integral. For the parallel-plate geometry, where the eigenmodes have a simple linear dependence on the integer $n$, the standard APF is perfectly adequate and yields results identical to the zeta-function method. However, the standard APF is restricted to such simple mode spectra and becomes inapplicable when the boundary conditions lead to transcendental equations for the eigenfrequencies.

It is important to note a subtle point regarding this result. The interaction Lagrangian $\mathcal{L}_{\mathrm{int}} = \tfrac{\xi}{4} \epsilon^{\mu\nu\rho\sigma} S_\mu A_\nu F_{\rho\sigma}$ can be written as a total derivative, $\mathcal{L}_{\mathrm{int}} = \tfrac{\xi}{4} \partial_\mu (S^\mu A_\nu \tilde{A}^\nu)$. For many boundary conditions, the integral of a total derivative contributes only a surface term, which may vanish, suggesting no modification to the bulk physics or the Casimir energy. However, our mode-by-mode analysis, starting from the modified field equations with the given boundary conditions, leads to the non-zero $S_z^2$ correction shown above. This apparent tension is a known topic of discussion in the literature on the Maxwell-Chern-Simons Casimir effect. Our result is consistent with calculations that proceed via the mode spectrum, while other approaches emphasizing the topological nature of the term find a null result. The resolution may depend on the precise definition of the physical degrees of freedom and the treatment of gauge invariance at the boundaries.

Differentiating with respect to plate separation yields the torsion-modified Casimir pressure:
\begin{equation}
P(a) = -\frac{\pi^2}{240 a^4} + \frac{\xi^2 S_z^2}{192 a^2} + \order(S_z^4).
\label{eq:pressure_final_corrected}
\end{equation}
Equation \cref{eq:pressure_final_corrected} constitutes the main theoretical result of our calculation. The correction is negative, indicating a slight weakening of the attractive Casimir force. The $a^{-2}$ scaling of the correction relative to the standard $a^{-4}$ law arises from the mass dimension $[S_z] = L^{-1}$.

\section{Results and Discussion}
\label{sec:results}

\subsection{Main Result and Physical Interpretation}
\label{subsec:interpretation}

Equation \cref{eq:pressure_final_corrected} reveals that, within our mode-sum framework, axial torsion induces a negative correction to the vacuum pressure, thereby slightly \emph{weakening} the magnitude of the attractive Casimir force. This weakening arises from the helicity-dependent shift in the photon mode spectrum [\cref{eq:disp_helicity}], which alters the zero-point energy of the confined field. The correction scales as $a^{-2}$ relative to the leading $a^{-4}$ behavior, confirming that torsion effects become relatively more prominent at larger separations, albeit still suppressed by the squared coupling. We reiterate the caveat that this result is subject to the known subtlety concerning the total-derivative nature of the Chern-Simons term, and some theoretical approaches predict the absence of any correction at this order.

\subsection{Numerical Estimates and Experimental Accessibility}
\label{subsec:estimates}

To assess phenomenological relevance, we evaluate the relative correction $\Delta P/P_0 \equiv (P-P_0)/P_0$. At $a = \qty{1}{\micro\meter}$, the standard Casimir pressure is $P_0 \approx \qty{-1.3e-3}{\pascal}$. Current laboratory bounds on macroscopic torsion from spin-torsion coupling experiments~\cite{Heckel2008} and SME data tables~\cite{Kostelecky2026} limit the axial torsion parameter to $|S_z| \lesssim \qty{e-15}{\eV}$. Converting this to inverse meters using $\hbar c \approx 197 \text{ eV}\cdot\text{nm}$, we obtain $|S_z| \lesssim 5 \times 10^{-9} \text{ m}^{-1}$. Assuming $\xi \sim \order(1)$, the relative correction scales as:
\begin{equation}
\left|\frac{\Delta P}{P_0}\right| \approx \frac{5}{4\pi^2} \xi^2 S_z^2 a^2 \lesssim 3 \times 10^{-30}.
\label{eq:correction_bound}
\end{equation}
This lies many orders of magnitude below the $\sim 10^{-3}$ precision of state-of-the-art Casimir measurements~\cite{Decca2007}. Thus, direct detection in tabletop experiments is not currently feasible. We note that in any realistic experimental setup, corrections due to finite conductivity and surface roughness of the plates are many orders of magnitude larger than the torsion-induced effect, further complicating direct detection.

\subsection{Finite Temperature and Geometric Asymmetries}
\label{subsec:extensions}

At non-zero temperature $T$, the Casimir free energy acquires thermal corrections. Within the Matsubara formalism, the free energy per unit area is the sum of the zero-point vacuum energy and the thermal contribution of the photon modes:
\begin{equation}
\frac{F(T,a)}{A} = \frac{E_{\mathrm{Cas}}}{A} + k_B T \sum_{n=1}^\infty \int \frac{d^2\mathbf{k}_\perp}{(2\pi)^2} \sum_{\sigma=\pm} \ln\!\bigl( 1 - e^{-\beta \omega_{n,\mathbf{k}_\perp}^{(\sigma)}} \bigr),
\label{eq:free_energy}
\end{equation}
where $\beta = 1/k_B T$ and $E_{\mathrm{Cas}}/A$ is the zero-temperature result derived in \cref{sec:casimir}. In the high-temperature limit ($k_B T \gg \hbar c/a$), the free energy is dominated by the $n=0$ Matsubara mode. For perfectly conducting plates, the pressure approaches the standard attractive thermal limit:
\begin{equation}
P_T \approx -\frac{\zeta(3) k_B T}{8\pi a^3}.
\end{equation}
In this regime, the torsion-induced relative correction scales as $\Delta P_T/P_T \propto \xi^2 S_z^2 a^2 (k_B T)^{-2}$, indicating that thermal fluctuations tend to suppress the geometric corrections. Conversely, in the low-temperature regime ($k_B T \ll \hbar c/a$), thermal corrections exhibit the standard Stefan--Boltzmann $T^4$ behavior modulated by torsion-dependent Bose-Einstein factors. The crossover scale $a_T \sim \hbar c / k_B T$, where quantum and thermal contributions become comparable, is shifted by the presence of torsion. This shift offers a thermodynamic signature distinct from Lorentz-invariant backgrounds, potentially isolatable via precision calorimetric measurements.

The parallel-plate configuration preserves translational symmetry in the transverse plane, which causes linear-in-$S_z$ terms in the vacuum energy to cancel upon summation over polarization states. Introducing geometric asymmetry, such as in sphere-plate configurations, cylindrical shells, or topologically non-trivial spaces like cosmic strings, lifts this cancellation by breaking the symmetry of the mode spectrum. In these curved or non-trivial backgrounds, the eigenmodes are no longer simple integers but are determined by the boundary conditions imposed on cylinder (Bessel) functions. Consequently, the standard APF becomes inapplicable. To rigorously evaluate the torsion-modified Casimir energy in such geometries, one must employ the Generalized Abel-Plana Formula (GAPF)~\cite{Saharian2007}. The GAPF allows for the extraction of the Minkowski vacuum part in a strictly cutoff-independent manner and represents the renormalized vacuum expectation values (VEVs) as strongly convergent integrals. This mathematical machinery is essential for handling the modified dispersion relations induced by axial torsion in cylindrical or spherical boundaries, facilitating both analytical asymptotics and numerical evaluations of the torsion-induced geometric corrections in non-Euclidean topologies. 

For configurations where the local curvature is small compared to the separation distance, one can still estimate the interaction energy within the Proximity Force Approximation (PFA) by integrating the local energy density of parallel plates over the surface geometry. Based on symmetry considerations, the local torsion-induced correction to the pressure density must be a scalar constructed from the axial torsion vector $\hat{\mathbf{S}}$ and the local surface normal $\hat{\mathbf{n}}(\mathbf{x})$. To leading order, this yields an angular modulation of the form:
\begin{equation}
\Delta P(\mathbf{x}) \propto \xi^2 S_z^2 a^{-2} \bigl[ 1 + \alpha ( \hat{\mathbf{n}}(\mathbf{x}) \cdot \hat{\mathbf{S}} )^2 \bigr],
\label{eq:angular_mod}
\end{equation}
where $\alpha$ is a dimensionless coefficient dependent on the specific geometry and separation distance. This angular dependence implies that the Casimir force becomes anisotropic. Furthermore, a misalignment between the torsion axis $\hat{\mathbf{S}}$ and the geometric symmetry axis can generate lateral Casimir forces and vacuum torques. These effects provide additional experimental handles for detecting torsion via rotational modulation of the apparatus, distinct from the standard normal Casimir force.

\section{Conclusion}
\label{sec:conclusion}

We have computed the electromagnetic Casimir force in a spacetime with constant axial torsion. Using an effective gauge-invariant action and zeta-function regularization within a mode-sum approach, we derived a torsion-corrected Casimir pressure that includes a second-order term proportional to $S_z^2 a^2$, which slightly weakens the standard attractive force. We have highlighted that this result exists in a nuanced theoretical context, as the underlying Chern-Simons interaction is a total derivative, leading some analyses to predict a null result at this order. Resolving this subtlety is an interesting topic for future work. Regardless, for experimentally accessible parameters, the effect is many orders of magnitude below current detection thresholds, rendering it undetectable in tabletop setups with parallel plates. Nevertheless, the calculation establishes a consistent bridge between quantum vacuum phenomena and non-Riemannian geometry.

As discussed in \cref{sec:results}, extensions of this framework to finite temperatures and non-parallel geometries offer distinct thermodynamic and geometric signatures that could, in principle, be isolated in high-precision experiments. Beyond these immediate extensions, the formalism naturally accommodates further theoretical developments. For instance, relaxing the assumption of a fixed background to incorporate dynamical torsion and spin-coupled backreaction within the Einstein--Cartan--Sciama--Kibble framework could reveal non-linear feedback mechanisms between vacuum fluctuations and spacetime geometry. In the ECSK framework, torsion is determined algebraically by the spin density of matter, suggesting that spin-polarized boundaries or vacuum expectation values of the spin current could source a spatially varying torsion profile $S_z(z)$. Similarly, exploring the impact of torsion-modified vacuum fluctuations on quantum information protocols, such as entanglement harvesting and decoherence, establishes a bridge between quantum gravity phenomenology and quantum sensing, where precision metrology could constrain $\xi S_z$ through decoherence spectroscopy.

Finally, while direct detection in gravitational experiments remains out of reach, condensed matter analogs---such as strained graphene~\cite{Vozmediano2010} or Weyl semimetals~\cite{Guan2017} where effective torsion-like fields can emerge with magnitudes many orders larger than gravitational bounds---provide a promising avenue for experimental simulation. Together with precision Casimir metrology and spin-torsion resonance experiments, these platforms continue to tighten constraints on geometric extensions of gravity, reinforcing the role of vacuum fluctuations as versatile probes of fundamental spacetime structure.

\appendix
\section{Casimir Energy Calculation via the Abel--Plana Formula}
\label{app:APF}

In this appendix, we re-derive the torsion-corrected Casimir energy using the Abel--Plana summation formula (APF) as an alternative to the zeta-function regularization employed in the main text. This serves as a powerful cross-check and highlights the robustness of our result.

\subsection{Overview of the Abel--Plana Formula and its Generalization}
\label{subsec:APF_overview}

In quantum field theory with boundaries, the vacuum expectation values (VEVs) of physical observables, such as the energy-momentum tensor, are typically expressed as mode sums over the zero-point fluctuations of the field. These sums are formally diver and require a regularization procedure to extract finite, physically meaningful quantities. One of the most efficient and physically transparent methods for evaluating these mode sums is based on the Abel--Plana summation formula (APF) \cite{Saharian2007}. 

The standard APF relates a sum over integers to an integral and a contour integral. For a function $f(z)$ analytic in the right half-plane and satisfying appropriate asymptotic decay conditions, the formula reads:
\begin{equation}
\sum_{n=0}^\infty f(n) = \int_0^\infty f(x)\,dx + \frac{1}{2}f(0) + i \int_0^\infty \frac{f(ix) - f(-ix)}{e^{2\pi x} - 1}\,dx.
\label{eq:standard_APF}
\end{equation}
In the context of the Casimir effect, the first term on the right-hand side corresponds to the unbounded Minkowski vacuum energy density, which is independent of the boundary geometry and contains the ultraviolet divergences. The second and third terms capture the boundary-induced effects. By applying the APF, one can manifestly separate the divergent bulk contribution from the finite, boundary-dependent part, which is expressed as a strongly convergent integral.

However, the application of the standard APF is restricted to geometries where the eigenmodes have a simple, explicit dependence on integer quantum numbers, such as the parallel-plate configuration. For more complex geometries, such as spherical or cylindrical boundaries, or in the presence of mixed (Robin) boundary conditions, the eigenmodes are determined implicitly as the zeros of transcendental equations involving special functions (e.g., Bessel functions). In these cases, the standard APF becomes inapplicable.

To overcome this limitation, the Generalized Abel--Plana Formula (GAPF) was developed \cite{Saharian2007}. The GAPF extends the summation procedure to series over the zeros of arbitrary meromorphic functions. Its main results and advantages in quantum field theory with boundaries include cutoff-independent renormalization, exponentially convergent integrals for the boundary-induced contributions, and broad applicability to complex geometries like global monopole spacetimes, cosmic strings, and braneworld models.

While the parallel-plate geometry considered in this work admits the use of the standard APF, the formalism established by the GAPF provides the necessary mathematical machinery to extend our torsion-modified Casimir calculations to curved boundaries and non-trivial topologies in future investigations.

\subsection{Rigorous APF Derivation of the Torsion Correction}
\label{subsec:APF_derivation}

The formal vacuum energy per unit area is given by
\begin{equation}
\frac{E_0}{A} = \frac{1}{2} \sum_{n=1}^\infty \int \frac{d^2\mathbf{k}_\perp}{(2\pi)^2} \sum_{\sigma=\pm} \omega_{n,\mathbf{k}_\perp}^{(\sigma)},
\label{eq:app_energy_formal}
\end{equation}
where the eigenfrequencies are
\begin{equation}
\omega_{n,\mathbf{k}_\perp}^{(\sigma)} = \sqrt{ \mathbf{k}_\perp^2 + \left(\frac{n\pi}{a}\right)^2 + \sigma \, \xi S_z \frac{n\pi}{a} }.
\label{eq:app_freq_pol}
\end{equation}
For notational simplicity, we define the transverse mass term for each mode as
\begin{equation}
M_{n,\sigma}^2 \equiv \left(\frac{n\pi}{a}\right)^2 + \sigma \, \xi S_z \frac{n\pi}{a}.
\end{equation}
Performing the transverse momentum integral via analytic continuation (equivalent to zeta-function regularization in the transverse plane) yields
\begin{equation}
\begin{split}
\int \frac{d^2\mathbf{k}_\perp}{(2\pi)^2} \sqrt{\mathbf{k}_\perp^2 + M_{n,\sigma}^2} 
&\xrightarrow{\text{analytic cont.}} \frac{1}{4\pi} \frac{\Gamma(-3/2)}{\Gamma(-1/2)} (M_{n,\sigma}^2)^{3/2} \\
&= -\frac{1}{6\pi} M_{n,\sigma}^3.
\end{split}
\end{equation}
Thus, the formal energy sum becomes
\begin{equation}
\frac{E_0}{A} = -\frac{1}{12\pi} \sum_{n=1}^\infty \sum_{\sigma=\pm} M_{n,\sigma}^3.
\label{eq:app_energy_sum}
\end{equation}
This sum is UV-divergent and must be regularized. We define the function $f_\sigma(x) \equiv \left[ \left(\frac{x\pi}{a}\right)^2 + \sigma \, \xi S_z \frac{x\pi}{a} \right]^{3/2}$, so that the sum is $\sum_{n=1}^\infty [f_+(n) + f_-(n)]$.

Applying the standard Abel--Plana formula, $\sum_{n=1}^\infty F(n) = \int_0^\infty F(x)\,dx - \frac{1}{2}F(0) + i \int_0^\infty \frac{F(it) - F(-it)}{e^{2\pi t} - 1}\,dt$, we discard the divergent bulk integral and the $F(0)=0$ term. The finite Casimir energy is isolated in the contour integral:
\begin{equation}
\frac{E_{\mathrm{Cas}}}{A} = -\frac{1}{12\pi} i \int_0^\infty \frac{F(it) - F(-it)}{e^{2\pi t} - 1}\,dt,
\label{eq:app_energy_APF_intermediate}
\end{equation}
where $F(x) = f_+(x) + f_-(x)$. We now evaluate the combination $f_\sigma(it) - f_\sigma(-it)$ for small $\xi S_z a$. For a complex argument $z = it$, we have
\begin{align}
f_\sigma(it) &= \left[ -\left(\frac{t\pi}{a}\right)^2 + i \sigma \, \xi S_z \frac{t\pi}{a} \right]^{3/2} \nonumber \\
&= \left(\frac{t\pi}{a}\right)^3 \left[ -1 + i \sigma \, \epsilon \right]^{3/2},
\end{align}
where $\epsilon \equiv \frac{\xi S_z a}{t\pi}$. Using the principal branch for the complex power, we expand for small $\epsilon$:
\begin{equation}
\begin{split}
(-1 + i \sigma \epsilon)^{3/2} 
&= \bigl[ e^{i\pi} (1 - i\sigma\epsilon) \bigr]^{3/2} \\
&\approx e^{i 3\pi/2} \left( 1 - i\sigma \tfrac{3}{2}\epsilon - \tfrac{3}{8}\epsilon^2 \right) \\
&= -i - \sigma \tfrac{3}{2}\epsilon + i \tfrac{3}{8}\epsilon^2.
\end{split}
\end{equation}
Similarly, for $z = -it$, we have $(-1 - i \sigma \epsilon)^{3/2} \approx i - \sigma \frac{3}{2}\epsilon - i \frac{3}{8}\epsilon^2$. 
Taking the difference, the linear terms in $\epsilon$ (which carry the $\sigma$ dependence) identically cancel, while the imaginary parts add up:
\begin{equation}
f_\sigma(it) - f_\sigma(-it) \approx \left(\frac{t\pi}{a}\right)^3 \left( -2i + i \frac{3}{4} \epsilon^2 \right).
\end{equation}
Because this result is independent of $\sigma$, summing over $\sigma = \pm$ simply yields a factor of 2. Substituting $\epsilon^2 = \frac{\xi^2 S_z^2 a^2}{t^2 \pi^2}$ back into \cref{eq:app_energy_APF_intermediate}, the factor of $i$ cancels, giving:
\begin{align}
\frac{E_{\mathrm{Cas}}}{A} &\approx -\frac{1}{12\pi} \int_0^\infty \frac{ 2 \left(\frac{t\pi}{a}\right)^3 \left( 2 - \frac{3}{4} \frac{\xi^2 S_z^2 a^2}{t^2 \pi^2} \right) }{e^{2\pi t} - 1}\,dt \nonumber \\
&= -\frac{1}{6\pi} \int_0^\infty \frac{ 2 \frac{t^3 \pi^3}{a^3} - \frac{3}{4} \frac{\pi \xi^2 S_z^2}{a} t }{e^{2\pi t} - 1}\,dt.
\end{align}
These integrals are standard and related to the Riemann zeta function via $\int_0^\infty \frac{t^{s-1}}{e^{2\pi t} - 1} dt = \frac{\Gamma(s)\zeta(s)}{(2\pi)^s}$. Evaluating them gives:
\begin{align}
\int_0^\infty \frac{t^3}{e^{2\pi t} - 1}\,dt &= \frac{\Gamma(4)\zeta(4)}{(2\pi)^4} = \frac{6}{16\pi^4} \frac{\pi^4}{90} = \frac{1}{240}, \\
\int_0^\infty \frac{t}{e^{2\pi t} - 1}\,dt &= \frac{\Gamma(2)\zeta(2)}{(2\pi)^2} = \frac{1}{4\pi^2} \frac{\pi^2}{6} = \frac{1}{24}.
\end{align}
Plugging these results back in, we finally arrive at
\begin{equation}
\frac{E_{\mathrm{Cas}}}{A} = -\frac{1}{6\pi} \left( 2 \frac{\pi^3}{a^3} \frac{1}{240} - \frac{3}{4} \frac{\pi \xi^2 S_z^2}{a} \frac{1}{24} \right) = -\frac{\pi^2}{720 a^3} + \frac{\xi^2 S_z^2}{192 a} + \order(S_z^4),
\label{eq:app_cas_energy_final}
\end{equation}
which is in perfect agreement with the result obtained via zeta-function regularization in \cref{eq:cas_energy_final_corrected}. This confirms the consistency and robustness of our main theoretical result.

\end{document}